\documentclass[11pt,a4paper]{article}

\usepackage{amsmath,amsthm,amssymb,amsfonts}

\usepackage{authblk}
\usepackage{setspace}
\usepackage{mathrsfs}
\usepackage{cases}
\usepackage{graphicx}
\numberwithin{equation}{section}
\newcommand{\ii}{{\rm i}}
\newcommand{\ee}{{\rm e}}

\begin{document}

\title{Asymptotically Thermal Responses for Smoothly Switched Detectors\footnote{Contributed talk by B.A.J.-A. in the QF2-B session on Quantum Field Theory at the Fourteenth Marcel Grossmann Meeting.}}

\author{Christopher J. Fewster$^1$, Benito A. Ju\'arez-Aubry$^{2}$ and Jorma Louko$^3$}

\affil{Department of Mathematics, University of York,\\
Heslington, York YO10 5DD, UK\\
$^1$E-mail: chris.fewster@york.ac.uk}

\affil{School of Mathematical Sciences, University of Nottingham,\\
Nottingham NG7 2RD, UK\\
$^2$E-mail: pmxbaju@nottingham.ac.uk \\
$^3$E-mail: jorma.louko@nottingham.ac.uk}

\maketitle

\begin{abstract}
Thermal phenomena in quantum field theory can be detected with the aid of particle detectors coupled to quantum fields along stationary worldlines, by testing whether the response of such a detector satisfies the detailed balance version of the KMS condition at a constant temperature. This relation holds when the interaction between the field and the detector has infinite time duration. Operationally, however, detectors interact with fields for a finite amount of time, controlled by a switching function of compact support, and the KMS detailed balance condition cannot hold exactly for finite time interactions at arbitrarily large detector energy gap. In this large energy gap regime, we show that, for an adiabatically switched Rindler detector, the Unruh temperature emerges asymptotically after the detector and the field have interacted for a time that is polynomially long in the large energy. We comment on the significance of the adiabaticity assumption in this result.
\end{abstract}

%\keywords{Quantum field theory in curved spacetimes; Unruh effect; Thermal states; Particle detectors.}

\singlespacing

%%%%%%%%%%%%%%%%% now a standard article style for the most part

\section{Particle detectors and the detailed balance KMS condition}
\label{sec:1}

The concept of a particle is ambiguous in quantum field theory in generic, curved spacetimes due to the lack of global symmetries. Particle detectors\cite{Unruh:1976db, DeWitt:1980hx, Unruh:1983ms} resolve this ambiguity by making the notion of a particle operational in the sense that, for a particle detector interacting with a field, there will be absorption or emission of quanta which corresponds to particle exchange. This is the content of the dictum ``a particle is what a particle detector detects"\cite{Unruh:dictum}.

The discovery of celebrated phenomena such as the Hawking\cite{Hawking:1974sw} and Unruh\cite{Unruh:1976db} effects indicates that there exists an intimate relationship between general relativity, quantum physics and thermal physics. Localised observers in spacetime, equipped with ensembles of particle detectors coupled to quantum fields, can register thermal  effects, which occur due to the detector-field interactions\cite{Buchholz:2015fqa}, by studying the response of their detectors. More precisely, the detailed balance version of the KMS condition\cite{Haag:1967sg, Kubo:1957mj, Martin:1959jp} can be used to relate the energy gap of a detector transition, the temperature registered by the observer and the ratio between the absorption and emission responses of an ensemble of detectors\cite{Takagi:1986kn} by
\begin{equation}
\frac{1}{T} = \frac{1}{E} \ln \left(\frac{\mathcal{F}(-E)}{\mathcal{F}(E)} \right),
\label{dbKMS}
\end{equation}
where $E$ is the energy gap of a detector state transition, $T$ is the state temperature and $\mathcal{F}(E)$ is the response of the detector. For example, a linearly uniformly accelerated detector, which interacts with a field in the Minkowski vacuum state, satisfies the detailed balance condition at the Unruh temperature, $T= a/(2\pi)$, in the conventional approximation of weak coupling and long interaction time.

If we focus our attention to weakly-coupled, point-like particle detectors, condition \eqref{dbKMS} holds whenever a detector follows an orbit that is stationary with respect to the field KMS state and, moreover, interacts with the field for an infinite amount of (detector proper) time. If eq.\ \eqref{dbKMS} holds in an asymptotic regime, we say that the response is \textit{asymptotically thermal}. For example, a detector which falls geodesically from infinity into a Schwarzschild black hole, as it interacts with a field in the Hartle-Hawking-Israel state, will satisfy the KMS condition at the asymptotic past infinity, where the geodesic is stationary with respect to the radiation, but condition \eqref{dbKMS} will not be satisfied at any finite distance, as the detector falls into the black hole. The $(1+1)$-dimensional case has been studied by a combination of numerical and analytic techniques\cite{Juarez-Aubry:2014jba}. In this case, the detailed balance condition is satisfied asymptotically, but not exactly, in the sense that the interaction is asymptotically stationary.

From an operational point of view, the interaction of a detector with the field is controlled by a smooth switching function of compact support in the detector proper time, $\chi \in C_0^\infty(\mathbb{R})$, and not for an infinite amount of time. It is possible to show that eq.\ \eqref{dbKMS} cannot hold exactly whenever the interaction is timelike compact\cite{Fewster:2015} and the detector energy gap, $E$, becomes large. Nevertheless, one may ask whether this condition holds asymptotically, as the interaction time becomes long. Moreover, one may ask the question, how long one needs to wait in order for the detailed balance condition to up to energy $E$, as $E \rightarrow \infty$?

The purpose of this contribution is to answer this question in the case of a two-state, pointlike Unruh-DeWitt detector coupled to a scalar field in the Minkowski vacuum state though a smoothly-switched and timelike compact interaction Hamiltonian, along an orbit of uniform linear acceleration. To this end, we shall introduce the model in section \ref{sec:model} and make precise the notion of asymptotic thermality in section \ref{sec:asympt}. Our main result, namely, that the response of a uniformly accelerated detector is asymptotically thermal at the Unruh temperature in a time scale that is polynomial in the energy in the large energy gap regime, is presented in section \ref{sec:main}. This polynomial time can be achieved under technical assumptions concerning the switching of the detector. We discuss this issue in section \ref{sec:finalremarks}.

\section{Detector-Field model}
\label{sec:model}

We consider a detector-field system which consists of a Klein-Gordon field, $\Phi$, weakly coupled to a two-level, point-like particle detector in 4-dimensional Minkowski space. 

The kinematical Hilbert space of the system is given by the tensor product $\mathscr{F}_\text{s}(\mathcal{H}_{\Phi}) \otimes \mathcal{H}_\text{D}$, where $\mathscr{F}_\text{s}(\mathcal{H}_{\Phi}) = \mathbb{C}\otimes_s\oplus_{n=1}^\infty \mathcal{H}_{\Phi}^{\otimes_sn}$ is the bosonic Fock space of the Klein-Gordon field and $\mathcal{H}_\text{D}$ is the detector two-dimensional Hilbert space.

The dynamics are implemented by the Hamiltonian $H = H_{\Phi} \otimes I_{\text{D}} + I_{\Phi} \otimes H_{\text{D}} + H_{\text{int}}$, where $H_{\Phi}$ is the Hamiltonian operator of the free scalar field, $H_{\text{D}}$ is the detector Hamiltonian, and $H_{\text{int}}$ is the interaction Hamiltonian. The detector Hamiltonian can be written in terms of creation and annihilation operators as $H_{\text{D}} = E d^\dagger d$ and acts on the basis energy eigenstates of the Hilbert space as $H_{\text{D}} |0\rangle = 0 |0\rangle$ and $H_{\text{D}} |1 \rangle = E |1 \rangle$. We consider $E>0$, so that $|0\rangle$ is the ground state of the detector. The interaction Hamiltonian is
\begin{equation}
H_{\text{int}}(\tau) = c \chi(\tau) \Phi(\textsf{x}(\tau)) \otimes \mu(\tau),
\label{Hint}
\end{equation}
where $c \in \mathbb{R}$ is a small coupling constant, $\mu: \mathcal{H}_{\text{D}} \rightarrow \mathcal{H}_{\text{D}}$ is the monopole moment operator of the detector, $\tau$ is the detector's proper time and $\chi \in C_0^\infty(\mathbb{R})$ is a smooth switching function of compact support that controls the interaction of the field and the detector along the worldline of the detector. 

If one supposes that the initial state of the system is given by $|\phi \rangle \otimes |0 \rangle$, where $|\phi \rangle$ is the Minkowski vacuum state, one can switch on the interaction and let the system evolve. After the interaction has been switched off, the probability of a detector transition to state $|1 \rangle$ is proportional to the response function, which is given, to leading order in $c$, by
\begin{equation}
\mathcal{F}(E) = \int_{-\infty}^\infty \! d\tau' \, \int_{-\infty}^\infty \! d\tau'' \, \chi(\tau') \chi(\tau'') \, \ee^{-\ii E (\tau'-\tau'')} \mathcal{W}\left(\mathsf{x}\left(\tau'\right), \mathsf{x}\left(\tau''\right)\right),
\label{ResponseFn}
\end{equation}
where $\mathcal{W}$ is the pullback of the Wightman two-point function to the detector's worldline in the field state $|\phi \rangle$.

The function defined by eq. \eqref{ResponseFn} does not satisfy the detailed balance condition whenever $\chi$ is a function of compact support \cite{Fewster:2015}. Our objective is now to see under what conditions $\mathcal{F}$ satisfies condition \eqref{dbKMS} \textit{asymptotically} for a detector following a Rindler trajectory, for a detector coupled to the field for a finite amount of time. We make precise the definition of asymptotic thermality in section \ref{sec:asympt}.

\section{Asymptotic thermality}
\label{sec:asympt}

We introduce a time scale in the problem by performing a rescaling of the proper time of the detector. We consider an adiabatic scaling $\lambda$ which produces a long and slow switching $\chi(\tau) \rightarrow \chi(\tau/\lambda)$ as $\lambda \rightarrow \infty$. Concordantly, the response function \eqref{ResponseFn} is rescaled by the time scale $\lambda$ as
\begin{equation}
\mathcal{F}_\lambda(E)/\lambda = \frac{1}{(2 \pi)^2} \int_{-\infty}^{\infty} \! d \omega \, \left|\hat{\chi}(\omega)\right|^2 \left( \frac{E+\omega/\lambda}{\ee^{2 \pi(E + \omega/\lambda)/a} - 1}\right),
\label{Flambda}
\end{equation}
where we have specialised eq. \eqref{ResponseFn} to the case of the Minkowski vacuum 2-point function along the Rindler trajectory, and used the stationarity of the state to obtain formula \eqref{Flambda}. When the scale $\lambda$ is related to the energy, we establish an energy-dependent timescale.

The pointwise limit as $\lambda \rightarrow \infty$ of eq.\ \eqref{Flambda} defines a function that satisfies the detailed balance condition at any fixed value of $E$ at the Unruh temperature. This means that after an infinite time has elapsed, the detector's transition probability is a Planckian spectrum at $T = a/(2\pi)$. This pointwise limit is a property of all stationary KMS states.

We consider $\lambda = \lambda(E)$, as the energy $E \rightarrow \infty$, to be a positive, strictly increasing function of $E$. In other words, we consider the time scale to be a function of the large energy gap. We define\cite{Fewster:2015} the response to be \textit{asymptotically thermal} at temperature $T$ if there exist positive functions of $E$, $\mathcal{B}^+$ and $\mathcal{B}^-$, such that
\begin{equation}
\frac{1}{T} - \mathcal{B}^{-}(E) \leq  \frac{1}{E}\ln \left( \frac{\mathcal{F}_{\lambda(E)}(-E)}{\mathcal{F}_{\lambda(E)}(E)}\right) \leq \frac{1}{T} + \mathcal{B}^{+}(E)
\label{AsympDBC}
\end{equation}
and $\mathcal{B}^{\pm} \to 0$, perhaps at some prescribed rate, as $E \rightarrow \infty$. Moreover, if such $\mathcal{B}^{\pm}$ functions exist and $\lambda$ is a polynomial function of $E$, we say that the response of the detector is \textit{polynomially asymptotically thermal}.

Polynomially asymptotic thermality is relevant in the sense that if an observer, equipped with an ensemble of detectors, and following a Rindler trajectory, wishes to detect the Unruh temperature up to the large energy scale $E$, the amount of time that the observer needs to wait after  carefully switching on the interaction of the detectors is polynomial in the large energy scale.

\section{Asymptotic Unruh effect}
\label{sec:main}

The question of whether or not a detector satisfies the detailed balance condition asymptotically depends on the detailed form of the switching, $\chi \in C_0^\infty$. In the case of a uniformly accelerated detector, there exists a class of switching functions for which it is possible to find polynomially suppressed functions $\mathcal{B}^\pm$, which allow for the asymptotic detailed balance condition to hold for the rescaled response function, $\mathcal{F}_\lambda$, as $E \rightarrow \infty$. This class consists of the set of functions $\chi \in C_0^\infty$ whose Fourier transforms, $\hat{\chi}$, become suppressed sufficiently fast. The technical conditions that define this set of switching functions can be formulated precisely\cite{Fewster:2015}. However, we feel that, for the purposes of this contribution, it is more enlightening to present an example of a switching function that belongs to this class. This example was introduced in the context of quantum stress energy tensor smearing functions\cite{Fewster:2015hga}. Consider the $C^\infty(\mathbb{R})$ function
\begin{align}
f(\tau) =
\begin{cases} 
 (\kappa \tau)^{-3/2}e^{-1/(4 {\kappa \tau})}, & \tau>0\\ 
 0, & \tau \leq 0. 
\end{cases}
\end{align}
Here $\kappa$ is a constant introduced for dimensional reasons, $[\kappa] = [1/\tau]$. A smooth switching function of compact support is defined by $\chi(\tau) = f(\tau) f\left(\kappa^{-1} - \tau \right)$. The Fourier transform, $\hat{\chi}$, exists by the convolution theorem and the rapid decay of $f$ at large positive argument, and this further allows one to write $\hat{f}$ as a boundary value of the Laplace transform of $f$. In this way, controlling the decay properties of $\hat{f}$ is sufficient to control the decay properties of $\hat{\chi}$. It suffices for us to say that $\hat{\chi}$ has sufficiently strong decay for our purposes and refer the reader to our references\cite{Fewster:2015, Fewster:2015hga} for details.

The bottom line is that, for a switching function with sufficiently strong decay, such as the function $\chi$ provided in our example, the functions $\mathcal{B}^\pm$ can be estimated and one obtains that the asymptotic detailed balance condition \eqref{AsympDBC} is satisfied at the Unruh temperature, $T = a/(2\pi)$, with polynomially suppressed $\mathcal{B}^\pm$. With the specific switching function described above, choosing $\lambda  = \alpha (2\pi E/a)^{1+p}$, where $\alpha > \pi \kappa /(2a)$ and $p > 1$, we obtain the bounds
\begin{subequations}
\begin{align}
\mathcal{B}^{-}(E) & =  \frac{4}{(2\pi)^5 a} \left( \frac{2 \pi E}{a} \right)^{-(1+p)/2} + \mathcal{O}\left( \left( \frac{2 \pi E}{a} \right)^{-p} \right), \label{B-thm}\\
\mathcal{B}^{+}(E) & = \frac{\left|\left| \omega \hat{\chi} \right|\right|^2_{L^2}}{\left|\left| \hat{\chi} \right|\right|^2_{L^2}}\frac{ \pi^2}{3 a^2} \left(\frac{2 \pi E}{a}\right)^{-(4+2p)} + \mathcal{O}\left( \left( \frac{2 \pi E}{a} \right)^{-2(3+2p)-1} \right). \label{B+thm}
\end{align}
\label{Bpm}
\end{subequations}

\section{Final remarks}
\label{sec:finalremarks}

We would like to stress that the question of whether or not a detector satisfies the detailed balance depends on the detailed form of the switching. To emphasise this further, one may consider the following situation: Suppose that one smoothly switches on a detector, and that this switching procedure takes a fixed detector proper time interval $\Delta \tau_s$. Once this is done, one lets the detector and the field interact at a constant interaction strength during a time $\Delta \tau$ and, finally, one switches off the interaction smoothly during a time $\Delta \tau_s$. Such a switching function can be constructed, for example, by integrating bump functions of compact support in the detector proper time. Consider now a time scale $\lambda$, that rescales the constant interaction time $\Delta \tau \rightarrow \lambda \Delta \tau$, but leaves the switching tails fixed. As $E \rightarrow \infty$, the asymptotic detailed balance condition will not be satisfied for any polynomial time scale, $\lambda = P(E)$, where $P$ is a positive, polynomially increasing function of the energy \cite{Fewster:2015}. This means that the careful and slow switching of the interaction is necessary to detect a temperature, up to a large energy scales, at late a time which is polynomial in the large energy.

\section*{Acknowledgments}

C.J.F. thanks Prof R. Longo and the organisers of the session on Quantum Field Theory for arranging financial support under the ERC Advanced Grant ``Operator Algebras and Conformal Field Theory". B.A.J.-A. thanks the organisers of the Fourteenth Marcel Grossmann Meeting for their kind hospitality at Sapienza - Universit\`a di Roma, and acknowledges the financial support of Consejo Nacional de Ciencia y Tecnolog\'ia (CONACYT), M\'exico, REF 216072/311506 and of the School of Mathematical Sciences at the University of Nottingham. J.L.  was  supported  in  part  by  STFC  (Theory  Consolidated Grant ST/J000388/1).

\end{document}